\documentclass[letter, reprint, pra, aps, superscriptaddress]{revtex4-1}

\def\ttitle{Ex Vacuo Atom Chip Bose-Einstein Condensate (BEC)}

\usepackage{graphicx}
\usepackage{array}
\usepackage[pdfpagelabels=true]{hyperref}
\usepackage[all]{nowidow}

\hypersetup{
  naturalnames=true,
  colorlinks=true,
  linkcolor=blue,
  pdfpagemode=UseNone,
  pdfstartview=FitH,
  pdftitle={\ttitle},
  pdfauthor={Matthew B. Squires},
  pdfsubject={Membrane-based vacuum system for rapid-prototype atom chips},
  pdfkeywords={Cold atoms, Atom chips},
}

\begin{document}

\def\AFRL{
  Air Force Research Laboratory, Space Vehicles Directorate, 
  Kirtland Air Force Base, NM
}
\def\SDLnm{
  Space Dynamics Laboratory, 
  Kirtland Air Force Base, NM
}
\def\SDLma{
  Space Dynamics Laboratory, 
  North Logan, UT 
}

\title{\ttitle}
\author{Matthew B. Squires}
\author{Spencer E. Olson}
\affiliation{\AFRL}
\author{Brian Kasch}
\affiliation{\AFRL}
\affiliation{\SDLnm}
\author{James A. Stickney} 
\affiliation{\SDLma}
\author{Christopher J. Erickson}
\author{Jonathan A. R. Crow}
\author{Evan J. Carlson}
\author{John H. Burke}
\affiliation{\AFRL}

\date{\today}

\begin{abstract}

Ex vacuo atom chips, used in conjunction with a custom thin walled vacuum
chamber, have enabled the rapid replacement of atom chips for magnetically trapped cold atom
experiments.  Atoms were trapped in $>2$~kHz magnetic traps created using high
power atom chips.  The thin walled vacuum chamber allowed the atoms to be trapped 
$\lesssim1$~mm from the atom chip conductors which were located
outside of the vacuum system. Placing the atom chip outside of the vacuum 
simplified the electrical connections and improved thermal management.  Using a 
multi-lead Z-wire chip design, a Bose-Einstein condensate was produced with an 
external atom chip. Vacuum and optical conditions were maintained while 
replacing the Z-wire chip with a newly designed cross-wire chip. The atom chips 
were exchanged and an initial magnetic trap was achieved in less 
than three hours.

\end{abstract}

\maketitle

\section{Introduction}

`Atom chips' are used to trap neutral atoms with magnetic fields generated using
wires precisely patterned on an insulating 
substrate\cite{art:AtomChipsReichel}. The benefits of using atom chips for 
trapping and controlling cold neutral atoms include precision magnetic field 
generation, reduced power consumption, low inductance for high speed response, and 
high gradient magnetic traps (e.g. $B'\gtrsim 20$~kG/cm). The high magnetic
field gradients of atom chips have typically been generated by trapping atoms
close ($\lesssim100~{\rm\mu m}$) to on-chip 
conductors~\cite{art:BECPortableAtomChipDu,
art:FermiDegeneracyOnAChipThywissen,
art:RFdoubleWellInterferometrySchmiedmayer, art:CombinedAtomChipZimmermann,
art:FastAtomChipBECHorikoshi}. Trapping atoms $\lesssim 100~{\rm\mu m}$ from the atom
chip wires involves installing the atom chip in an ultra-high vacuum (UHV)
environment, which requires careful cleaning and bakeout at elevated
temperatures. Furthermore, placing atom-chip conductors inside the vacuum
requires careful thermal management, as well as UHV-compatible electrical
feedthroughs.

\begin{figure}[!ht]
  \centering
  a.)\includegraphics[width = 0.95 \columnwidth]{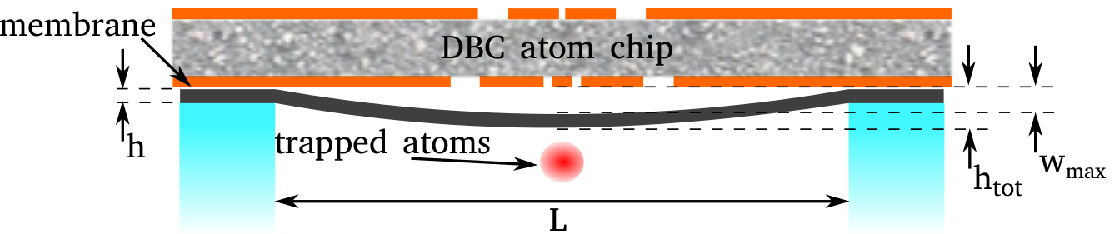} \\
  b.)\includegraphics[width = 0.95 \columnwidth]{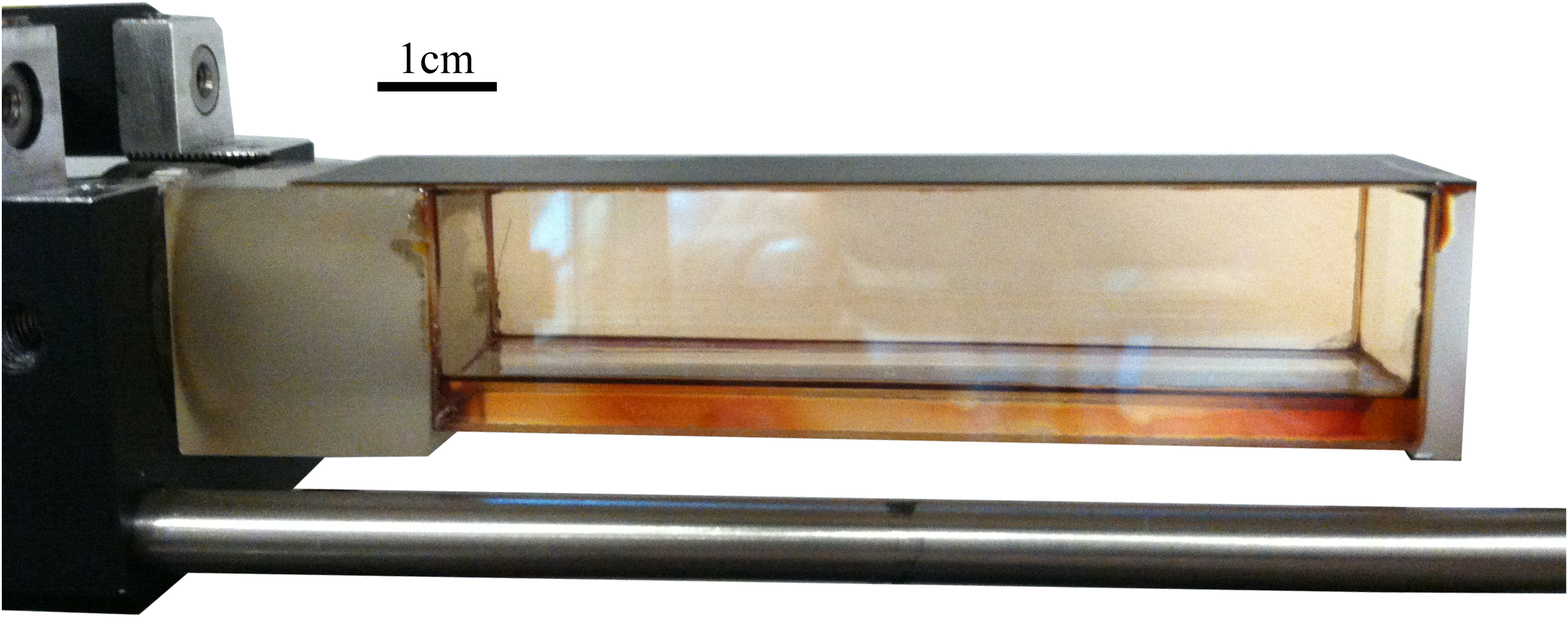} \\
  \vspace{0.2in}
  c.) \includegraphics[width = 0.9 \columnwidth]{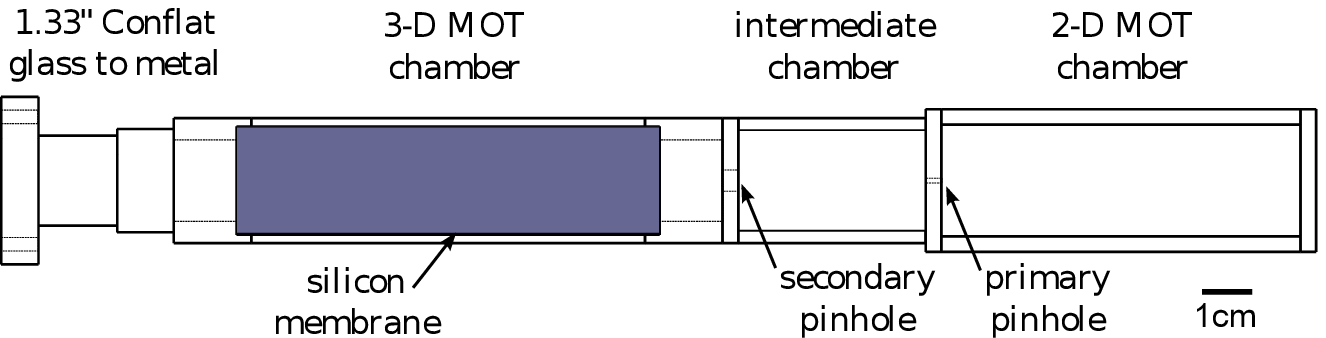}
  \caption{
    (Color online)
    a.) Schematic drawing of the membrane vacuum wall, atom chip, and trapped
    atoms. The distance from the atom chip wires to the atoms was the thickness
    of the membrane plus the bowing of the membrane.
    b.) Photograph of an epoxied vacuum cell with a 350$~\mu$m thick silicon
    membrane. The vacuum side of th\hyphenation{thatshouldnot}e silicon was 
polished and silver coated.
    c.) Schematic top view of epoxied chamber.
  }
  \label{fig:membrane}
\end{figure}

In this paper atom chips were placed outside of the vacuum envelope and were
successfully used to trap laser cooled atoms. This was made possible by the 
combination of high power atom chips \cite{art:DBCAtomChipsSquires} in close 
proximity to a thin monocrytalline silicon membrane which formed one wall of a custom vacuum 
chamber. The thin membrane enabled the trapping of laser-cooled atoms 
$<1$~mm from the atom chip (see Fig.~\ref{fig:membrane}a). Tight atom-chip 
magnetic traps (i.e. trap frequencies $> 2$~kHz) were still possible, while external mounting 
of the chips enabled rapid and simple switching of atom chips without impacting 
the UHV environment. Production of a Bose-Einstein condensate (BEC) using ex 
vacuo atom chips, as well as trapping atoms in precision atom-chip magnetic 
potentials \cite{art:TunableTrapsStickney} were demonstrated. In comparison to 
other atom-chip systems \cite{art:ChipBECHansch, art:MicroElectroMagnetsPrentiss, 
art:FermiDegeneracyOnAChipThywissen, art:BECInterferometerOnAChipCornell,
art:TransportableBECFarkas}, the atom chips were switched multiple times
without breaking vacuum.  Less than three hours was required to switch the atom
chip and to achieve a magnetic trap with BEC compatible conditions.
Additionally, the external mounting greatly simplified 
electrical connections and thermal management.

\section{Construction}
\label{sec:construction}

Two vacuum chamber designs are presented in this section. The two vacuum 
chambers were similar in design, but with some key differences
such as silicon thickness (discussed in Sec.~\ref{sec:membrane_design}). The 
chamber designs are presented to demonstrate the ability vary various 
construction details while maintaining the same functionality.  

Both chambers used dual magneto-optical traps (MOTs) where a 2-D+
MOT~\cite{art:TransportableBECFarkas} loaded a 3-D
mirror MOT. A mirror MOT was used to allow the MOT to form in close
proximity to the atom chip~\cite{art:WireTrapForNeutralsHansch}, which assisted
in the loading of the magnetic trap after laser cooling. The mirror was
formed by depositing silver on the vacuum side of the silicon membrane. After laser
cooling, the atoms were magnetically trapped using an ex vacuo atom chip 
mounted above the 3-D MOT chamber.

The chambers were constructed from silicon and borosilicate glass 
joined with a UHV-compatible epoxy (Epo-tek 353ND)
\cite{art:AppsOfIntegratedMicrtrapsHansch}. This method has been reliably used 
for constructing custom UHV chambers utilizing components that would be
otherwise difficult to join using common glass blowing methods (e.g. 
double-sided anti-reflection coatings or dissimilar materials) or standard UHV
components.

\begin{figure}[ht]
  \centering
  \includegraphics[width = 0.99 \columnwidth]{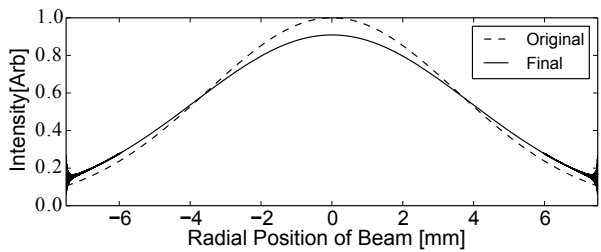}
  \caption{
    \label{fig:membrane_optical_front_5mm}
    Relative intensity change of a collimated beam after reflecting off a 
    bowed membrane at 5~mm from the mirrored
    surface. The solid line is a simulation intensity profile prior
    to reflection off the bowed membrane and the dashed line is the intensity
    profile after reflection and 5~mm of propagation, which is approximately where
    the MOT was formed. 
  }
\end{figure}

The first chamber was made using separate 2-D and \mbox{3-D} MOT cells 
joined on opposite faces of a Conflat cube. This design was similar to the 
configuration in Ref.
\cite{phdthesis:squires}, except that one of the long sides of the 3-D MOT cell
was sealed with a thin silicon membrane (see Fig. \ref{fig:membrane}b).
A $40~\ell/{\rm s}$ ion pump and titanium sublimation pump were connected to
one of the remaining ports. The exit aperture of the 2-D MOT was $1.5$~mm in
diameter. The silicon membrane was 150~$\mu$m thick and the deformation of
the silicon could be seen when the chamber was evacuated. The calculated focal
length of the deformed silicon was approximately $-44$~cm. As shown in
Fig.~\ref{fig:membrane_optical_front_5mm}, the resulting intensity change was
negligible within the MOT range. In this configuration the 3-D MOT loaded
$\sim 5 \times 10^8$ atoms.

\begin{figure}[ht]
 \centering
 a.) \includegraphics[width = 0.9 \columnwidth]{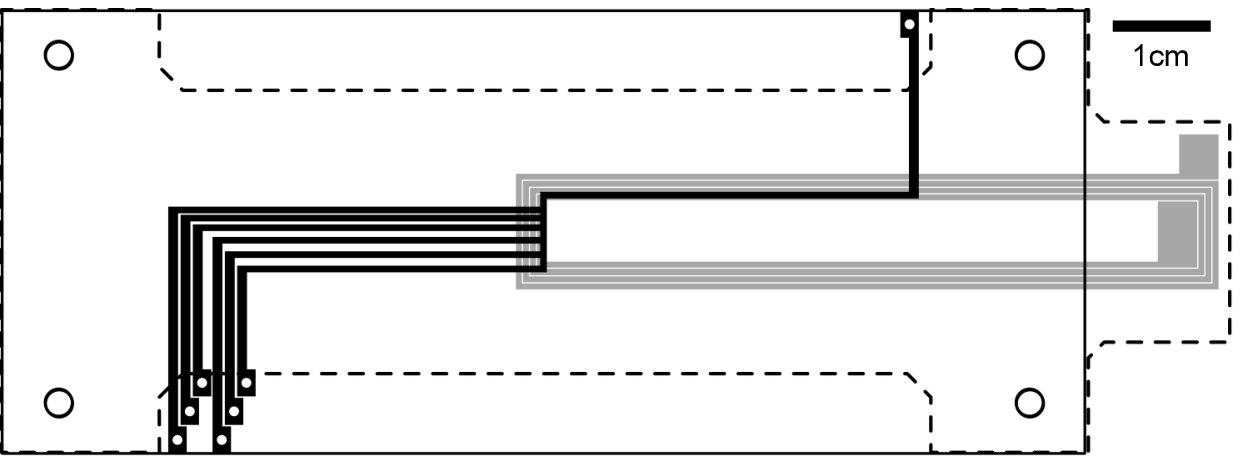}
 b.) \includegraphics[width = 0.9 \columnwidth]{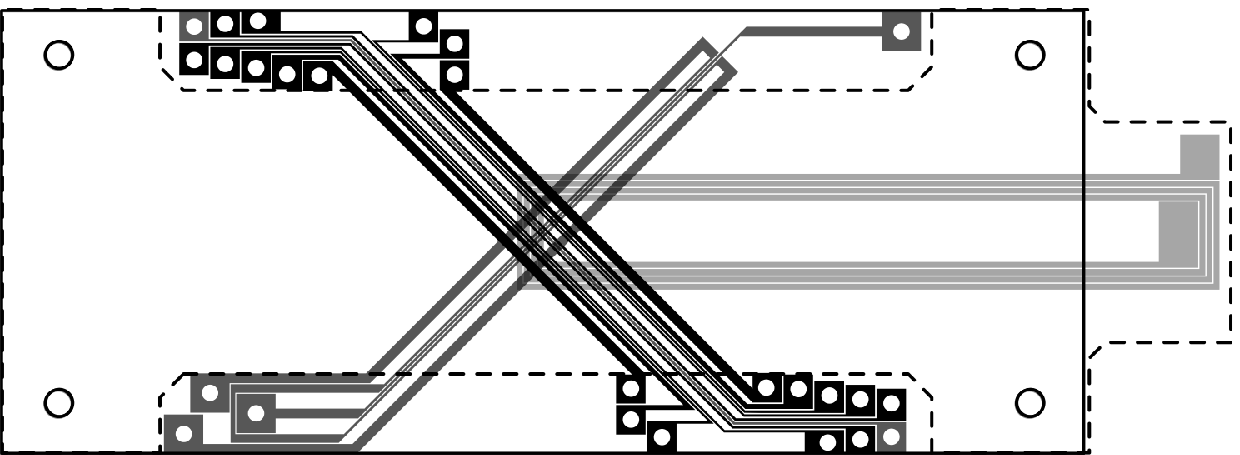}
 \caption{
 Overlay of the magnetic trapping chips and the U-wire chips. Each chip is
 a double-sided direct bonded copper (DBC) atom chip. The magnetic trapping 
conductors are shown in
 darker shading and the chip outline and mounting holes are shown as solid
 lines. The U-wire chip conductors are shown as light gray with the chip
 outline shown as a dashed line. a.) The multi-lead Z-wire for producing BEC.
 The wire traces are placed facing downward, toward the silicon membrane. b.)
 The top layer (black) of the cross wire trap chip consists of wire pairs used
 to make a controllable polynomial magnetic
 potential~\cite{art:TunableTrapsStickney}. The wire pattern on the front
 side of the chip (dark gray) was used for making a four wire waveguide. The
 center wires on both sides (dark gray) were used for magnetic trapping and BEC
 generation.}
 \label{fig:AtomChipDesigns}
\end{figure}

A magnetic trap was realized using this cell with multiple Z-wire iterations.
Fig.~\ref{fig:AtomChipDesigns}a illustrates the final Z-wire configuration.
During one of the atom chip changes, the silicon cracked along
a crystal boundary. The cracked silicon was repaired with VacSeal, however the
magnetic trap lifetime was degraded and a second chamber was built as
a replacement.

The second chamber was similar to the first chamber except for the following
changes: 1) the entire chamber was epoxied together with a single
1.33'' Conflat at the end of the 3-D MOT chamber; 2) the exit aperture of the 
2-D
  MOT was 0.5~mm; 3) the silicon membrane was approximately 350~$\mu$m thick; 
and 4)
  an intermediate chamber was epoxied between the 2-D and 3-D MOT chambers (see
  Fig. \ref{fig:membrane}c). The intermediate chamber was approximately 25~mm
  square and filled with SAES ST 707 non-evaporable getter strips for
  passive pumping of residue gases released by the rubidium dispensers (SAES RB
  AMD). The exit hole of the intermediate chamber was 2~mm diameter. The
  double pinhole design was implemented to improve differential pumping between
  the chambers. 
There was minimal deflection of the thicker silicon membrane. The MOT
number was $\sim 5 \times 10^8$ atoms. Because of the similar MOT 
numbers with two different deformations, it did not appear that the small deformation of the 
silicon noticeably affected the MOT number or temperature.

The atom chip traces were laser etched and the chip outline was cut using
a pulsed fiber laser milling system (Oxford Laser Systems) from
commercially available direct bonded copper (DBC) 
substrates~\cite{art:DBCAtomChipsSquires}. After laser
etching, the resistance between traces was 5--50~$\Omega$. At this resistance, 
the chip wires were suitably isolated and the residual
conductivity was due to copper slag from the laser etching process. A 
1--30~s chemical etch in either HCl/H$_2$O$_2$ or cupric
chlorate~\cite{art:CopperEtchingCakir} removed any residual copper. The
final resistance between isolated copper areas was $>5$~M$\Omega$.

Mounting and alignment holes were laser cut after laser etching the wire traces 
without removing the DBC substrate from the mill. These mounting holes were assumed
to be aligned with the chip traces within the tolerance of the laser mill ($\sim 1\; \mu$m).
The alignment holes were used to mount the atom chips relative to the vacuum chamber
and allowed for precision alignment between the various atom chip layers
(e.g. the U-wire chip and the BEC chips).

\section{Membrane Design and Testing}
\label{sec:membrane_design}

\begin{figure}[!ht]
  \centering
  \includegraphics[width = 0.95 \columnwidth]{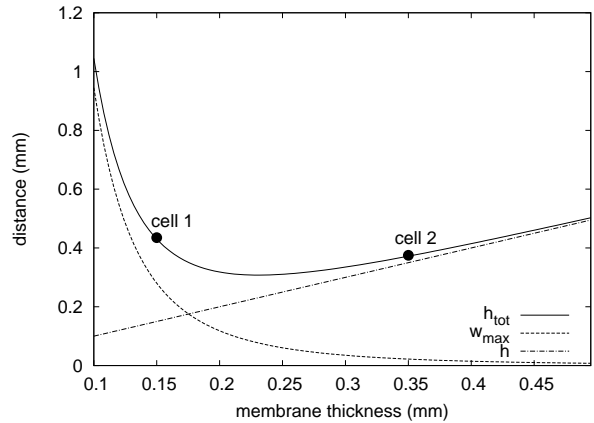}
  \caption{
    Plot of calculated total distance from the atom chip conductors to the atom
    trapping location as a function of membrane thickness assuming a silicon
    membrane, $L = 1.5$~cm, $p = 1.01 \times 10^ 5$Pa, and that the membrane is
    longer than it is wide, such that $c_1 =  0.0284$.
  }
  \label{fig:OptimalMembraneThickness}
\end{figure}

Trapping atoms as closely as possible to the atom chip conductors is critical for
generating tight magnetic traps due to the $1/r^2$ nature of magnetic field
gradients. As seen in Fig.~\ref{fig:membrane}a, the effective distance from an
atom chip to the vacuum is $h_{tot} = h + w_{max}$, where $h$ is the thickness
of the membrane and $w_{max}$ is the maximum displacement due to the bowing of
the thin membrane. If the
membrane is too thin, the distance to the atoms increases due to the bowing of
the membrane (see Fig.~\ref{fig:OptimalMembraneThickness}). Conversely, if the membrane is thick, there is little bowing but
the distance to the atoms increases due to the thickness of the membrane.  

The maximum deflection at the center of a rectangular plate with fixed edges and
uniformly applied pressure is
\begin{equation}
 w_{max} = c_1 \frac{p L^4}{E\, h^3}
\label{eqn:MaxPlateDeflection}
\end{equation}
where $c_1$ is a numerical factor determined by the shape of the membrane, $L$
is the width of the cell, $E$ is the Young's modulus of the material, $p$ is the
pressure, and $h$ is the thickness of the
membrane~\cite{book:MechanicsOfMaterialsHearn}. The optimal membrane thickness
that minimizes the distance to the atoms is
\begin{equation} 
 h_{opt} = L \left(\frac{3 c_1 p}{E}\right)^{1/4}.
 \label{eqn:OptimalMembraneThickness}
\end{equation}
In considering which materials to use, those with nearly
matched coefficients of thermal expansion (CTE) were chosen to minimize the 
stress of the vacuum chamber during bakeout. 
Young's modulus has little effect on optimal membrane thickness because of the
$E^{-1/4}$ scaling dependence. Thus, optimizing $E$ has minimal benefit. Silicon 
was
used for the membrane and borosilicate glass for the rest of the chamber 
because of the nearly matched CTEs and polished silicon wafers are available in 
a wide 
variety of thicknesses.

\begin{table}
 \begin{tabular}{|c||c|c|c|}
\hline
Material      & CTE ($10^{-6}/$K) & E (GPa) & Strength (MPa) \\ \hline  \hline
fused quartz  & 0.55      & 72        & 48 \\ \hline
diamond (CVD) & 1.0       & 1,050     & 750  \\ \hline
borosilicate  & 3.25      & 64     & 35--100  \\ \hline
Si            & 3.55        & 112--165  & 7,000  \\ \hline
SiC           & 4.0       & 405       & 3,440 \\ \hline 
W             & 4.5       & 400--410  & 1,510 \\ \hline
AlN           & 4.8       & 308       & 400 \\ \hline
sapphire      & 5.3       & 335       & 400 \\ \hline
WC            & 6         & 450--650  & 345\\ \hline
S.S. 302      & 15--18      & 200       & 520 \\ \hline
\end{tabular} 
\label{tab:materialParameters}
\caption{
    The coefficients of thermal expansion, Young's moduli, and strengths of
    relevant materials. The listed strengths of the materials are the lesser
    of their yield or ultimate strengths, depending on the brittleness of the
    material.
}
\end{table}

The plausibility of using thin a silicon membrane as a vacuum wall was tested
using a glass test cell with a $~15 \times 45$~mm opening that was sealed by
a 150~$\mu$m thick silicon membrane. This test membrane survived repeated 
pump/vent cycles and a vented cross-country laboratory move.

Separate burst testing was performed to determine the strength margin of the
membrane and whether the silicon crystal orientation had any effect on the burst
strength. A $12 \times 25$~mm pocket was milled in an aluminum block and sanded
flat. A variety of silicon membranes were epoxied to the aluminum block which
was placed in a pressurized tube to simulate the evacuation of a vacuum
chamber. The tube was over-pressurized to determine the strength of the
membrane above atmospheric pressure. The majority of membranes ruptured at
greater than 5 atmospheres of over-pressure with no apparent
dependence on crystal orientation.

\section{Results/Discussion}
\label{sec:ResultsDiscussion}

The implementation of laser cooling, magnetic trap, and RF evaporation were
similar to other atom chip experiments~\cite{art:ChipBECHansch,
art:BECPortableAtomChipDu, art:TransportableBECFarkas} with the exception of a 
few key experimental details highlighted here. Evanescently coupled 
polarization maintaining fiber splitters were used to divide the laser into 
multiple beams and to combine the cooling and repump lasers. A 
$2\times3$($2\times4$) splitter was used for the 2-D(3-D) MOT.
The 2-D MOT used two $10 \times 25$~mm elliptical beams and a 
15~mm circular pusher beam; the 3-D MOT employed four 15~mm circular beams.

The MOT quadrupole field was generated with a U-wire atom 
chip~\cite{art:WireTrapForNeutralsHansch}. Four loops on each side of
the U-wire chip were connected by a soldered via (see
Fig.~\ref{fig:AtomChipDesigns}). The U-wire chip was mounted on top of the
BEC atom chip using common mounting holes. This allowed for a consistent
alignment between the MOT and the other magnetic field generating chips and 
facilitated the rapid achievement of a magnetic trap after switching atom chips.

The MOT was loaded for $\sim15$~s at 8.0~mm from the silicon surface. After
loading, the atoms were cooled, compressed, and moved closer to the membrane 
within 80~ms during the compressed MOT (CMOT) stage. The magnetic fields were
then ramped off for molasses cooling. The U-wire and bias magnetic field coils
were controllably ramped off in 1~ms to ensure the position of the CMOT did
not significantly move prior to molasses cooling. The lasers were then further
detuned for 6 ms of molasses cooling. The cooling laser was then
shuttered. The atoms were then optically pumped into the $\left| F=2, m_F
= 2 \right>$ state. After optical pumping there were approximately $300 \times
10^6$ atoms at 15~$\mu$K.

These atoms were initially captured in a Z-wire magnetic trap using the largest
Z (see Fig.~\ref{fig:AtomChipDesigns}a) with 47.9~A in the Z-wire and $B_x
= 18.7$~G, which formed the initial trap at 4.6~mm from the silicon surface.
The largest Z was used for the initial trap because the depth of the 
trap depends on the trap height to Z-wire width ratio, i.e., beyond a maximum 
distance from the wires there is no longer a trap.  For two parallel wires with 
current in the same direction, no trap was formed when the height was greater than 
$\sqrt{3}$ width.

After the initial trap, the cloud was compressed by increasing 
the bias field and trapping in successively smaller Z-wires. The Z-wire electrical
configuration was changed by linearly switching the current between adjacent Z-wire 
leads to maintain a constant current in common lead of the
Z-wire. The contribution to the B$_y$ field from the Z-wire depended on the
width of the Z-wire. The B$_y$ bias field was adjusted during the
compression to maintain a constant bottom magnetic field during the Z-wire lead
switching. The final trap used the next to smallest Z-wire with external bias
fields $(35.0, 30.0, 0.0)$~G and $I_z = 39$~A for calculated trap frequencies
of $f = (50, 300, 300)$~Hz and a trap at 1.25~mm from the silicon surface. The
atoms were not ramped closer at this stage because of surface losses due to the
proximity of silicon membrane.  

The atoms were then cooled via RF evaporation. As the cloud cooled, it became
smaller which allowed the trap to be compressed and ramped closer to the chip
without additional surface losses. The RF sweep and compression sweeps
followed a $x(t)^E$ functional form. The first sweep ramped the RF frequency
from 50~MHz -- 6~MHz and the B$_x$ field was ramped from 35 -- 75~G with $E
= 0.5$ in 6~s. The final calculated trap parameters were $f
= (350,2000,2000)$~Hz at 0.5~mm from the silicon surface at $I_{Z} = $~39~A and
$B = (75.0,32.0,0.0)$~G. After the first RF sweep the trap was not compressed
further.

A second RF sweep ramped the RF frequency from 6~MHz--4.9~MHz in 1~s with an
exponent factor $E = 0.4$.  The BEC transition was detected by a sharp 
change in the effective area of the atom 
cloud~\cite{art:ImagesEvaporationBEC:Close}.  The transition temperature was 
150~nK with $\sim 2\times10^4$ atoms.

After first achieving BEC in the multi-Z compression trap, the multi-Z atom
chip was replaced by the the atom chip shown in 
Fig.~\ref{fig:AtomChipDesigns}b. The central wires on each side of the 
atom chip were connected in series.  The field of those wires plus an 
additional bias field were used to create the magnetic trap. Unlike the Z-wire 
design, there was no height:width ratio that limited the maximum trapping 
distance. As configured here, the cross-wire trap had a non-zero magnetic field 
bottom and was used for evaporative cooling and BEC production. The 
function of the other wires is described in 
Ref.~\cite{art:TunableTrapsStickney}.

The new atom chip was installed in approximately two hours. Most of that time
involved careful routing of the 26 electrical connections to the chip
(see Fig.~\ref{fig:AtomChipDesigns}b). The new atom chip was attached to the
chamber using the same mounting holes as the Z-wire atom chip. 

A MOT with nominal number and temperature was obtained within minutes of
installing the new atom chip assembly. A magnetic trap was achieved in less
than 30~minutes after installation. The new atom chip experiments are
described in Ref.~\cite{art:TunableTrapsStickney}. These experiments do not 
require BEC but rather atoms evaporatively cooled to $1-4~\mu$K. Evaporation 
using the new atom chip was as efficient as evaporation in the Z-wire atom 
chip.  

Since the first install, several wiring changes have been made with the new atom
chip configuration. These changes have required the atom chip to be removed and
reinstalled. Additionally, a BEC has been made with the cross-wire trap. Each
time the atom chip has been installed, nominal operating conditions have been
reestablished in less than 30~minutes with little optimization.

This paper has described the design and building of an atom chip system where ex 
vacuo atom chips are placed next to a thin membrane forming one wall of the
vacuum system. The thin membrane coupled with a high power DBC atom chip enables
tight magnetic traps compatible with BEC production. This setup
greatly simplifies the rapid replacement of atom chips. 
Magnetic traps and BECs have been produced with two different atom chip
configurations with minimal downtime between
experiments.

\section{Acknowledgements}
We gratefully acknowledge Benjamin Stuhl and Rudolph Kohn for their careful reading of this manuscript.

\bibliography{MembraneCellForAtomChips}

\end{document}